\documentclass[pss]{wiley2sp} 
\usepackage{multirow}
\usepackage{url}

\begin{document}

\title{Accuracy of Quantum Monte Carlo Methods for Point Defects in Solids}

\titlerunning{Testing QMC on Defects in Solids}

\author{%
  William D. Parker\textsuperscript{\textsf{\bfseries 1}},
  John W. Wilkins\textsuperscript{\textsf{\bfseries 1}},
  Richard G. Hennig\textsuperscript{\Ast\textsf{\bfseries 1,2}},
  }

\authorrunning{Parker et al.}

\mail{e-mail
  \textsf{rhennig@cornell.edu}, Phone: +01-607-2546429}

\institute{%
  \textsuperscript{1}\,Department of Physics, The Ohio State University, 191 W. Woodruff Ave., Columbus, Ohio 43210, USA\\
  \textsuperscript{2}\,Department of Materials Science and
  Engineering, Cornell University, Ithaca, New York 14853, USA}

\received{XXXX, revised XXXX, accepted XXXX} 
\published{XXXX} 

\pacs{ } 

\abstract{Quantum Monte Carlo approaches such as the diffusion Monte Carlo (DMC) method are among the most accurate many-body methods for extended systems.  Their scaling makes them well suited for defect calculations in solids.  We review the various approximations needed for DMC calculations of solids and the results of previous DMC calculations for point defects in solids.  Finally, we present estimates of how approximations affect the accuracy of calculations for self-interstitial formation energies in silicon and predict DMC values of $4.4(1)$, $5.1(1)$ and $4.7(1)$~eV for the X, T and H interstitial defects, respectively, in a 16(+1)-atom supercell.}

\maketitle

\section{Introduction}

Point defects, such as vacancies, interstitials and anti-site defects, are the only thermodynamically stable defects at finite temperatures~\cite{Tilley08}.  The infinite slope of the entropy of mixing at infinitesimally small defect concentrations results in an infinite driving force for defect formation.  As a result, at small defect concentrations, the entropy of mixing always overcomes the enthalpy of defect formations.  In addition to being present in equilibrium, point defects often control the kinetics of materials, such as diffusion and phase transformations and are important for materials processing.  The presence of point defects in materials can fundamentally alter the electronic and mechanical properties of a material.  This makes point defects technologically important for applications such as doping of semiconductors~\cite{Fahey89,Eaglesham94}, solid solution hardening of alloys~\cite{Pike97,Zhu99}, controlling the transition temperature for shape-memory alloys~\cite{Otsuka99} and the microstructural stabilization of two-phase superalloys.  
 
However, the properties of defects, such as their structures and formation energies, are difficult to measure in some materials due to their small sizes, low concentrations, lack of suitable radioactive isotopes, {\it etc.}  Quantum mechanical first-principles or {\it ab initio} theories make predictions to fill in the gaps left by experiment~\cite{Pelaz09}.

The most widely used method for the calculation of defect properties in solids is density functional theory (DFT).  DFT replaces explicit many-body electron interactions with quasiparticles interacting via a mean-field potential, {\it i.e.} the exchange-correlation potential, which is a functional of the electron density~\cite{Jones89}.  A universally true exchange-correlation functional is unknown, and DFT calculations employ various approximate functionals, either based on a model system or an empirical fit.  The most commonly used functionals are based on DMC simulations~\cite{Ceperley80} for the uniform electron gas at different densities, {\it e.g.}, the local density approximation (LDA)~\cite{Perdew81,PW92} and gradient expansions, {\it e.g.}, the generalized gradient approximation (GGA)~\cite{PW91,PBE,AM05,WC,PBESol}.  These local and semi-local functionals suffer from a significant self-interaction error reflected in the variable accuracy of their predictions for defect formation energies, charge transition levels and band gaps~\cite{Heyd05,Nieminen09}.  Another class of functionals, called hybrid functionals, include a fraction of exact exchange to improve their accuracy~\cite{Becke93,Heyd03}.

The seemingly simple system of Si self-interstitials exemplifies the varied accuracy of different density functionals and many-body methods. The diffusion and thermodynamics of silicon self-interstitial defects dominates the doping and subsequent annealing processes of crystalline silicon for electronics applications~\cite{Eaglesham94,Richie04,Du07}.  The mechanism of self-diffusion in silicon is still under debate.  Open questions~\cite{Vaidyanathan07} include: (1) are the interstitial atoms the prime mediators of self-diffusion, (2) what is the specific mechanism by which the interstitials operate, and (3) what is the value of the interstitial formation energy?  Quantum mechanical methods are well suited to determine defect formation energies. LDA, GGA and hybrid functionals predict formation energies for these defects ranging from about $2$ to $4.5$~eV~\cite{Batista06}.  Quasiparticle methods such as the $GW$~approximation reduce the self-interaction error in DFT and are expected to improve the accuracy of the interstitial formation energies.  Recent $G_{0}W_{0}$ calculations~\cite{Rinke09} predict formation energies of about $4.5$~eV in close agreement with HSE hybrid functionals~\cite{Batista06} and previous DMC calculations~\cite{Leung99,Batista06}.  Quantum Monte Carlo methods provide an alternative to DFT and a benchmark for defect formation energies~\cite{Foulkes01,Needs06}.

In this paper, we review the approximations that are made in diffusion Monte Carlo (DMC) calculations for solids and estimate how these approximations affect the accuracy of point defect calculations, using the Si self-interstitial defects as an example.  Section~\ref{sec:qmc} describes the quantum Monte Carlo method and its approximations.  Section~\ref{sec:review} reviews previous quantum Monte Carlo calculations for defects in solids, and Section~\ref{sec:results} discusses the results of our calculations for interstitials in silicon and the accuracy of the various approximations.

\section{Quantum Monte Carlo method}
\label{sec:qmc}

Quantum Monte Carlo (QMC) methods are among the most accurate electronic structure methods available and, in principle, have the potential to outperform current computational methods in both accuracy and cost for extended systems.  QMC methods scale as $O(N^3)$ with system size and can handle large systems.  At the present time, calculations for as many as 1,000 electrons on 1,000 processors make effective use of available computational resources~\cite{Batista06}.  Current work is under way to develop algorithms that extend the system size accessible by QMC methods to petascale computers~\cite{Esler08}.

Continuum electronic structure calculations primarily use two QMC methods~\cite{Foulkes01}: the simpler variational Monte Carlo (VMC) and the more sophisticated diffusion Monte Carlo (DMC).  In VMC, a Monte Carlo method evaluates the many-dimensional integral to calculate quantum mechanical expectation values.  Accuracy of the results depends crucially on the quality of the trial wave function which is controlled by the functional form of the wave function and the optimization of the wave functions parameters~\cite{Umrigar07}.  DMC removes most of the error in the trial wave function by stochastically projecting out the ground state using an integral form of the imaginary-time Schr\"odinger equation.

One of the most accurate forms of trial wave functions for quantum Monte Carlo applications to problems in electronic structure is a sum of Slater determinants of single-particle orbitals multiplied by a Jastrow factor and modified by a backflow transformation:
\begin{displaymath}
  \Psi(r_n) = \mathrm{e}^{J(r_n, R_m)} \, \sum c_i \, \mathrm{CSF}_i(x_n).
\end{displaymath}
The Jastrow factor $J$ typically consists of a low order polynomial and a plane-wave expansion in electron coordinates $r_n$ and nuclear coordinates $R_m$ that efficiently describe the dynamic correlations between electrons and nuclei.  Static (near-degeneracy) correlations are described by a sum of Slater determinants.  Symmetry-adapted linear combinations of Slater determinants, so-called configuration state functions (CSF), reduce the number of determinant parameters $c_i$.  For extended systems, the lack of size consistency for a finite sum of CSF's makes this form of trial wave functions impractical, and a single determinant is used instead. Finally, the backflow transformation $r_n \rightarrow x_n$ allows the nodes of the trial wave function to be moved which can efficiently reduce the fixed-node error~\cite{Lopez-Rios06}.  Since the backflow transformed coordinate of an electron $x_n$ depends on the coordinates of all other electrons, the Sherman-Morrison formula used to efficiently update the Slater determinant does not apply, increasing the scaling of QMC to $O(N^4)$.  If a finite cutoff for the backflow transformation is used, the Sherman-Morrison-Woodbury formula~\cite{Golub96} applies and the scaling is reduced to $O(N^3)$.

Optimization of the many-body trial wave function is crucial because accurate trial wave functions reduce statistical and systematic errors in both VMC and DMC.  Much effort has been spent on developing improved methods for optimizing many-body wave functions, and this continues to be the subject of ongoing research.  Energy and variance minimization methods can effectively optimize the wave function parameters in VMC calculations~\cite{Umrigar07,Umrigar88}.  Recently developed energy optimization methods enable the efficient optimization of CSF coefficients and orbital parameters in addition to the Jastrow parameters for small molecular systems, eliminating the dependence of the results on the input trial wave function~\cite{Umrigar07}.

VMC and DMC contain two categories of approximation to make the many-electron solution tractable: {\em controlled} approximations, whose errors can be made arbitrarily small through adjustable parameters, and {\em uncontrolled} approximations, whose errors are unknown exactly.  The controlled approximations include the finite DMC time step, the finite number of many-electron configurations that represent the DMC wave function, the basis set approximation, {\it e.g.}, spline or plane-wave representation, for the single-particle orbitals such of of the trial wave function and the finite-sized simulation cell.  The uncontrolled approximations include the fixed-node approximation which constraints the nodes of the wave function in DMC to be the same as the ones for the trial wave function, the replacement of the core electrons around each atom with a pseudopotential to represent the core-valence electronic interaction and the locality approximation that uses the trial wave function to project the nonlocal angular momentum components of the pseudopotential.

\subsection{Controlled approximations}

\paragraph{Time step}
Diffusion Monte Carlo is based on the transformation of the time-dependent Schr\"odinger equation into an imaginary-time diffusion equation with a source-sink term.   The propagation of the $3N$-dimensional electron configurations (walkers)  that sample the wave function requires a finite imaginary time step which introduces an error in the resulting energy~\cite{Anderson75,Umrigar93}.

Controlling the time step error is simply a matter of performing calculations for a range of time steps either to determine when the total energy or defect formation energy reaches the required accuracy or to perform an extrapolation to a zero time step using a low order polynomial fit of the energy as a function of time step.  Smaller time steps, however, require a larger total number of steps to sample sufficiently the probability space.  Thus, the optimal time step should be small enough to add no significant error to the average while large enough to keep the total number of Monte Carlo steps manageable.  In addition, the more accurate the trial wave function is the smaller the error due to the time-step will be~\cite{Umrigar93}.

\paragraph{Configuration population}
In DMC, a finite number of electron configurations represent the many-body wave function.  These configurations are the time-independent Schr\"odinger equation's analogues to particles in the diffusion equation and have also been called psips~\cite{Anderson75} and walkers~\cite{Foulkes01}.  To improve the efficiency of sampling the many-body wave function, the number of configurations is allowed to fluctuated from time step to time step in DMC using a branching step.  However, the total number of configurations needs to be controlled to avoid the configuration population to diverge or vanish~\cite{Umrigar93}.  This population control introduces a bias in the energy.  In practice where tested~\cite{Alfe04b}, few hundreds of configurations are sufficient to reduce the population control bias in the DMC total energy below the statistical uncertainty.

The VMC and DMC calculations parallelize easily over walkers.  After an initial decorrelation run, the propagation of a larger number of walkers is computationally equivalent to performing more time steps.  The variance of the total energy scales like
\begin{displaymath}
 \sigma_E^2 \propto \frac{\tau_\mathrm{corr}}{N_\mathrm{conf}\, N_\mathrm{step}}
\end{displaymath}
where $N_\mathrm{conf}$ denotes the number of walkers, $N_\mathrm{step}$ the number of time steps and $\tau_\mathrm{corr}$ the auto correlation time.


\paragraph{Basis set}
A sum of basis functions with coefficients represents the single-particle orbitals in the Slater determinant.  A DFT calculation usually determines these coefficients.  Plane waves provide a convenient basis for calculations of extended systems since they form an orthogonal basis that systematically improves with increasing number of plane waves that span the simulation cell.  Increasing the number of plane waves until the total energy converges within an acceptable threshold in DFT creates a basis set that has presumably the same accuracy in QMC.

Since the plane wave basis functions are extended throughout the simulation cell, the evaluation of an orbital at a given position requires a sum over all plane waves.  Furthermore, the number of plane waves is proportional to the volume of the simulation cell.  The computational cost of orbital evaluation can significantly be reduced by using a local basis, such as B-splines, which replaces the sum over plane waves with a sum over a small number of local basis functions.  The resulting polynomial approximation reduces the computational cost of orbital evaluation at a single point from the number of plane waves (hundreds to thousands depending on the basis set) to the number of non-zero polynomials (64 for cubic splines)~\cite{Alfe04}.  The wavelength of the highest frequency plane wave sets the resolution of the splines.  Thus, the most important quantity to control in the basis set approximation is the size of the basis set.

\paragraph{Simulation cell}
Simulation cells with periodic boundary conditions are ideally suited to describe an infinite solid but result in undesirable finite-size errors that need correction.  There are three types of finite-size errors.  First, the single-particle finite-size error arises from the choice of a single $k$-point in the single-particle Bloch orbitals of the trial wave function.  Second, the many-body finite size error arises from the non-physical self-image interactions between electrons in neighboring cells.  Third, the defect creates a strain field that results in an additional finite size error for small simulation cells.

The single-particle finite size error is greatly reduced by averaging DMC calculations for single-particle orbitals at different $k$-points that sample the first Brillouin zone of the simulation cell, so-called twist-averaging~\cite{Lin01} Alternatively, the single-particle finite size error can also be estimated from the DFT energy difference between a calculation with a dense $k$-point mesh and one with the same single $k$-point chosen for the orbitals of the QMC wave function.

For the many-body finite size error, several methods aim to correct the fictitious periodic correlations between electrons in different simulation cells.  The first approach, the model periodic Coulomb (MPC) interaction~\cite{Williamson97}, revises the Ewald method~\cite{Ewald21} to account for the periodicity of the electrons by restoring the Coulomb interaction within the simulation cell and using the Ewald interaction to evaluate the Hartree energy.  The second approach is based on the random phase approximation for long wave lengths.  The resulting first-order, finite-size-correction term for both the kinetic and potential energies can be estimated from the electronic structure factor~\cite{Chiesa06}.  The third approach estimates the many-body finite-size error from the energy difference between DFT calculations using a finite-sized and an infinite-sized model exchange-correlation functional~\cite{Kwee08}.   This approach relies on the exchange-correlation functional being a reasonable description of the system, whereas the other two approaches (MPC and structure factor) do not have this restriction.  The MPC and structure factor corrections are fundamentally related and often result in similar energy corrections~\cite{Drummond08}.

The defect strain finite size error, can be estimated at the DFT level using extrapolations of large simulation cells.  Also since QMC force calculations are expensive and still under development~\cite{Badinski10}, QMC calculations for extended systems typically start with DFT-relaxed structures.  Energy changes due to small errors in the ionic position as well as thermal disorder are expected to be quite small because of the quadratic nature of the minima and will largely cancel when taking energy differences for the defect energies.

\subsection{Uncontrolled approximations}

\paragraph{Fixed-node approximation}
The Monte Carlo algorithm requires a probability distribution, which is non-negative everywhere, but fermions, such as electrons, are antisymmetric under exchange, and therefore any wave function of two or more fermions has regions of positive and negative value.  For quantum Monte Carlo to take the wave function as the probability distribution, Anderson~\cite{Anderson75} fixed the zeros or nodes of the wave function and took the absolute value of the wave function as the probability distribution.  If the trial wave function has the nodes of the ground state, then DMC projects out the ground state.  However, if the nodes differ from the ground state, then DMC finds the closest ground state of the system within the inexact nodal surface imposed by the fixed-node condition.  This inexact solution has an energy higher than that of the ground state.

Three methods estimate the size of the fixed node approximation: (1) In the Slater-Jastrow form of the wave function, the single-particle orbitals in the Slater determinant set the zeroes of the trial wave function.  Since these orbitals come from DFT calculations, varying the exchange-correlation functional in DFT changes the trial wave function nodes and provides an estimate of the size of the fixed-node error.  (2) L\'opez R\'\i os {\it et al.}~\cite{Lopez-Rios06} applied backflow to the nodes by modifying the interparticle distances, enhancing electron-electron repulsion and electron-nucleus attraction.  The expense of the method has thus far limited its application in the literature to studies of second and third-row atoms, the water dimer and the 1D and 2D electron gases.  (3) Because the eigenfunction of the Hamiltonian has zero variance in DMC, a linear extrapolation from the variances of calculations with and without backflow to zero variance estimates the energy of the exact ground state of the Hamiltonian.

\paragraph{Pseudopotential}
Valence electrons play the most significant roles in determining a composite system's properties.  The core electrons remain close to the nucleus and are largely inert.  The separation of valence and core electron energy scales allows the use of a pseudopotential to describe the core-valence interaction without explicitly simulating the core electrons.  However, there is often no clear boundary between core and valence electrons, and the core-valence interaction is more complicated than a simple potential can describe.  Nonetheless, the computational demands of explicitly simulating the core electrons and the practical success of calculations with pseudopotentials in reproducing experimental values promote their continued use in QMC.  Nearly all solid-state and many molecular QMC calculations to date rely on pseudopotentials to reduce the number of electrons and the time requirement of simulating the core-electron energy scales.

Comparing DMC energies using pseudopotentials constructed with different energy methods (DFT and Hartree-Fock[HF]) provides an estimate of the error incurred by the pseudopotential approximation.  Additionally, the difference between density functional pseudopotential and all-electron energies estimates the size of the error introduced by the pseudopotential and is used as a correction term.

\paragraph{Pseudopotential locality}
DMC projects out the ground state of a trial wave function but does not produce a wave function, only a distribution of point-like configurations.  However, the pseudopotential contains separate potentials (or channels) for different angular-momenta of electrons.  One channel, identified as local, does not require the wave function to evaluate, but the nonlocal channels require an angular integration to evaluate, and such an integration requires a wave function.  Mit\'a\v s {\it et al.}~\cite{Mitas91} introduced use of the trial wave function to evaluate the nonlocal components requiring integration.  This locality approximation has an error that varies in sign.  While there are no good estimates of the magnitude of this error, Casula~\cite{Casula06} developed a lattice-based technique that makes the total energy using a nonlocal potential an upper bound on the ground-state energy.  Pozzo and Alf\`e~\cite{Pozzo08} found that, in magnesium and magnesium hydride, the errors of the locality approximation and the lattice-regularized method are comparably small, but the lattice method requires a much smaller time step ($0.05$~Ha$^{-1}$ vs. $1.00$~Ha$^{-1}$ in Mg and $0.01$~Ha$^{-1}$ vs. $0.05$~Ha$^{-1}$ in MgH$_2$) to achieve the same energy.  Thus, they chose the nonlocal approximation.

While all-electron calculations would, in principle, make the pseudopotential and locality errors controllable, in practice, the increase in number of electrons, required variational parameters and variance of the local energy makes such calculations currently impractical for anything but small systems and light elements~\cite{Esler10}.

\begin{figure*}[htb]
 \caption{DMC, $GW$~and DFT energies (in eV) for neutral defects in three materials.  DMC and experimental values have an estimated uncertainty indicated by numbers in parenthesis.  For the diamond vacancy, DFT-LDA and DMC include a $0.36$~eV Jahn-Teller relaxation energy.  LDA relaxation produced the structures and transition path so the DMC value for migration energy is an upper bound on the true value.  The Schottky energy in MgO is the energy to form a cation-anion vacancy pair.  DFT-LDA produces a range from $6$-$7$~eV depending on the representation of the orbitals and treatment of the core electrons.  DMC using a plane-wave basis and pseudopotentials results in a value on the upper end of the experimental range.  For Si interstitial defects, DFT values of the formation energy range from $2$~eV below up to the DMC values, depending on the exchange-correlation functional(LDA, GGA[PBE] or hybrid[HSE]), and the $GW$~values lie within the two-standard-deviation confidence level of DMC.}
  \label{tab:defect_review}
  \begin{center}
    \begin{tabular}{| l  c  c |  c |  c  c  c |  c | c | c | c |}
     \hline
       \multicolumn{3}{| l |}{} 	   		          & Energy    & \multicolumn{3}{c |}{DFT} & $GW$ & DMC             & Exp. & Ref.\\
       \multicolumn{3}{| l |}{} 	   		          & type      & LDA & GGA & Hybrid       &    &                 & & \\
      \hline                                                                                                            
      C  & \multicolumn{2}{l |}{diamond vacancy} & Formation & 6.98& 7.51& -             & -  & 5.96(34)       & - & \cite{hood:076403,PhysRevB.71.035206} \\
         &\multicolumn{2}{l |}{}& Migration & 2.83& -   & -             & -  & 4.40(36)  & 2.3(3) & \\
      \hline                  
      MgO&\multicolumn{2}{l |}{Schottky defect}& Formation  & 5.97, 6.99, 6.684& - & - & - & 7.50(53) & 5 - 7 & \cite{PhysRevB.71.220101,PhysRevB.76.184103}\\
      \hline
         & self-                          & X             &           & 3.31& 3.64& 4.69          &4.40& 5.0(2), 4.94(5)  & - &  \cite{Leung99,Batista06}\\
      Si & interstitial                    & T             & Formation & 3.43& 3.76& 4.95          &4.51& 5.5(2), 5.13(5)  & - & \\
         & defect& H             &	      & 3.31& 3.84& 4.80          &4.46& 4.7(2), 5.05(5)  & - & \\
      \hline
   \end{tabular}
  \end{center}
\end{figure*}

\section{Review of previous DMC defect calculations}
\label{sec:review}

To date, there have been DMC calculations for defects in three materials: the vacancy in diamond, the Schottky defect in MgO and the self-interstitials in Si.  

\subsection{Diamond vacancy}
Diamond's high electron and hole mobility and its tolerance to high temperatures and radiation make it a technologically important semiconductor material.  Diffusion in diamond is dominated by vacancy diffusion~\cite{Bernholc88}, and the vacancy is also associated with radiation damage~\cite{Collins09}.  Table~\ref{tab:defect_review} shows the range of vacancy formation and migration energies calculated by LDA~\cite{PhysRevB.71.035206} and DMC~\cite{hood:076403}.  DMC used structures from LDA relaxation and single-particle orbitals employing a Gaussian basis.  LDA pseudopotential removed core electrons.   The DMC calculations predict a lower formation energy than LDA.  The DMC value for the migration energy is an upper bound on the actual number since the structures have not been relaxed in DMC.  Furthermore, DMC estimates the experimentally observed dipole transition and provides an upper bound on the migration energy.~\cite{hood:076403} The GR1 optical transition is not a transition between one-electron states but between spin states $^{1}$E and $^{1}$T$_{2}$.   DMC calculates a  transition energy of $1.5(3)$~eV from $^{1}$E to $^{1}$T$_{2}$, close to the experimentally observed value of $1.673$~eV.  LDA cannot distinguish these states.  For the cohesive energy, DMC predicts a value of $7.346(6)$~eV in excellent agreement with the experimental result of $7.371(5)$~eV while LDA overbinds and yields $8.61$~eV. 

\subsection{MgO Schottky defect}
MgO is an important test material for understanding oxides.  Its rock-salt crystal structure is simple, making it useful for computational study.  Schottky defects are one of the main types of defects present after exposure to radiation, according to classical molecular dynamics simulations~\cite{Uberuaga05}.  Table~\ref{tab:defect_review} shows that DMC predicts a Schottky defect formation energy in MgO at the upper end of the range of experimental values~\cite{PhysRevB.71.220101}.

\subsection{Si interstitial defects}
Table~\ref{tab:defect_review} shows that DFT and DMC differ by up to $2$~eV in their predictions of the formation energies of these defects~\cite{Leung99,Batista06}.  We compare the DMC values with our results including tests on the QMC approximations in Section~\ref{sec:results}.

\begin{figure*}[htb]
  \caption{DMC Si defect formation energies.  Varying parameters and improved
    methods produce values for each defect that lie within two standard deviations
    of each other although the energetic ordering of the defects varies.  All
    calculations use DFT-LDA to produce the orbitals in the Slater determinant.}
  \label{tab:si_interstitial}
  \begin{center}
    \begin{tabular}{| c c c | c c c c | c | c | c | c |}
      \hline
      \multicolumn{3}{|c|}{Defect formation energy (eV)}& \multicolumn{4}{|c|}{Jastrow factor}&Pseudo-&Plane-wave&Finite-size& Reference\\
      \multicolumn{3}{|c|}{}& \multicolumn{3}{|c}{$n$-body terms} & plane-                &potential&cutoff energy&correction       & \\
       X     & T     & H&  e-e & e-n & e-e-n                   & waves    &method &(eV)& & \\
     \hline                                                                               
      \multicolumn{3}{|l|}{Slater-Jastrow} & & & & & & & &  \\
      5.0(2) &5.5(2) &4.7(2) & 16  & 16  & 0                       & 0            & LDA     &245 & DFT k-pt.&~\cite{Leung99}\\
      4.94(5)&5.13(5)&5.05(5)& 5   & 5   & 0                       & 0            & HF      &435 & DFT k-pt. &~\cite{Batista06}     \\
      4.9(1)&5.2(1)&4.9(1)& 8   & 8   & 3                       & 19           & DF      &1088& DFT k-pt.+struc. fac. & (this work) \\[0.5em]
      \multicolumn{3}{|l|}{Slater-Jastrow backflow} & & & & & & & & \\
      4.5(1)&5.1(1)&4.7(1)& 8   & 8   & 3                       & 19           & DF      &1088& DFT k-pt.+struc. fac. & (this work)  \\[0.5em]
      \multicolumn{3}{|l|}{Extrapolation} & & & & & & & & \\
      {\bf 4.4(1)}& {\bf 5.1(1) }& {\bf 4.7(1)}& 8   & 8   & 3                       & 19           & DF      &1088& DFT k-pt.+struc. fac. & (this work)  \\
      \hline
    \end{tabular}
  \end{center}
\end{figure*}

\section{Results}
\label{sec:results}

We specifically test the time-step, pseudopotential and fixed-node approximations for the formation energies of three silicon self-interstitial defects, the split-$\langle 110\rangle$~interstitial (X), the tetrahedral interstitial (T) and the hexagonal interstitial (H).  The QMC calculations are performed using the {\sc casino}~\cite{CASINO} code.  Density functional calculations in this work used the {\sc Quantum ESPRESSO}~\cite{QE-2009} and {\sc WIEN}2k~\cite{Wien2k} codes.  The defect structures are identical to those of Batista {\it et al.} ~\cite{Batista06}.  The orbitals of the trial wave function come from DFT calculations using the LDA exchange-correlation functional.  The plane-wave basis set with a cutoff energy of $1,088$~eV ($60$~Ha) converges the DFT total energies to $1$~meV.  A 7$\times$7$\times$7 Monkhorst-Pack $k$-point mesh centered at the L-point (0.5,0.5,0.5) converges the DFT total energy to $1$~meV.  A population of $1,280$ walkers ensured that the error introduced by the population control is negligible small.  Due to the computational cost of backflow, we perform the simulations for a supercell of 16(+1) atoms and estimate the finite-size corrections using the structure factor method~\cite{Chiesa06}.  The final corrected DMC energies for the X, T and H defects are shown in the bottom line of Table~\ref{tab:si_interstitial}.

\begin{figure}[htb]
  \includegraphics*[width=8cm]{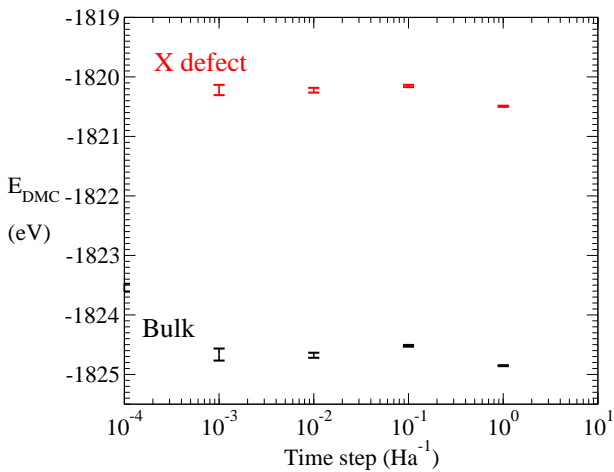}
  \caption{DMC total energies with varying (imaginary) time steps for bulk silicon and the X defect.  The error due to the finite-sized time step is smaller than the statistical uncertainty in the total energies for values from $10^{-3}$ to $1$~Ha$^{-1}$.  Note that these energies include no finite-size or pseudopotential corrections and thus differ in value from those in Figure~\ref{fig:dmc_extrapolation}.}
  \label{fig:timestep}
\end{figure}

\subsection{Time step}

Figure~\ref{fig:timestep} shows the total energies of bulk silicon and the X defect as a function of time step in DMC.  A time step of $0.01$~Ha$^{-1}$ reduces the time step error to within the statistical uncertainty of the DMC total energy.

\subsection{Pseudopotential}
In our calculations, a Dirac-Fock (DF) pseudopotential represents the core electrons for each silicon atom~\cite{Trail05a,Trail05b,CASINOpp}.  To estimate the error introduced by the pseudopotential, we compare the defect formation energies in DFT using this pseudopotential with all-electron DFT calculations using the linearized augmented plane-wave method~\cite{Wien2k}.  This comparison gives corrections of $0.083$, $-0.168$ and $0.054$~eV for the H, T and X defects respectively.

\begin{figure}[htb]
  \includegraphics*[width=8cm]{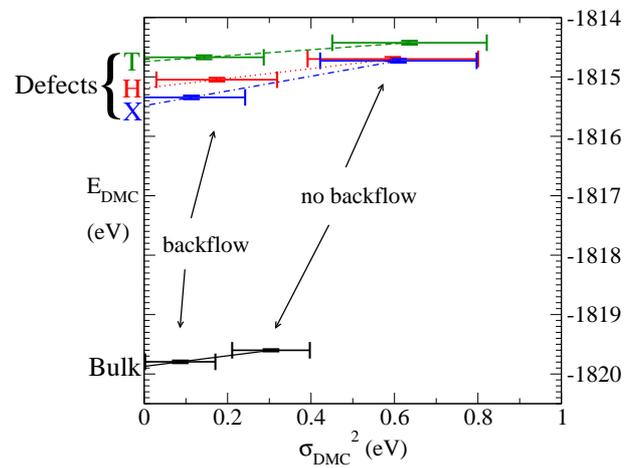}
  \caption{DMC total energies calculated with and without backflow transformation and linearly extrapolated to zero variance.  The backflow transformation reduces the total energies by 0.19(3), 0.62(5), 0.25(5) and 0.35(5)~eV for bulk (solid line) and the X (dash-dot line), T (dashed line) and H (dotted line) defects respectively.  The extrapolation to zero variance only slightly reduces the total energies by about 0.1~eV.  The reduction of the total energy from the backflow transformation indicates the size of the errors due to the fixed-node and pseudopotential locality approximation.}
  \label{fig:dmc_extrapolation}
\end{figure}

\subsection{Fixed-node approximation}
The fixed-node approximation is the main source of error of diffusion Monte-Carlo calculations.  To estimate the size of the error introduced by the fixed-node approximation, we perform the calculations using the backflow transformation which allows the nodes of the trial wave function to be moved and reduces the fixed-node error~\cite{Lopez-Rios06}.  We estimate the error due to the fixed-node approximation by performing calculations with and without the backflow transformation and by extrapolating the resulting defect formation energies to zero variance.  

Applying the backflow transformation to electron coordinates using polynomials of electron-electron, electron-nucleus and electron-electron-nucleus separations, we include polynomial terms to eighth-order for each spin type in electron-electron separation, to sixth-order in electron-nucleus separation and to third-order for each spin type in electron-electron-nucleus separation.  Figure~\ref{fig:dmc_extrapolation} shows the linear extrapolation of the DMC energies for the Slater-Jastrow and Slater-Jastrow-backflow trial wave function to zero variance.  The total energy decrease for the bulk and interstitial cells due to the backflow transformation ranges from 0.20(5) to 0.62(5)~eV.   The backflow transformation results in a significantly improved nodal surface of the trial wave function which is reflected in the reduced the variance of the local energy.  

Table~\ref{tab:si_interstitial} list the Si interstitials formation energies in DMC for the Slater-Jastrow, Slater-Jastrow-backflow wave function and the extrapolation.  Applying the backflow transformation reduces the formation energies for the X, T and H interstitial by 0.42(5), 0.05(5) and 0.15(5) eV, respectively.  The linear extrapolation provides a simple estimate of the remaining fixed node error.  The extrapolation lowers the interstitial formation energy by a negligible amount of 0.06(10), 0.00(10), 0.07(10)~eV for the X, T and H interstitial, respectively.  The resulting Si interstitial formation energies are 4.4(1), 5.1(1) and 4.7(1)~eV for the X, T and H interstitial, respectively, in close agreement with recent $G_{0}W_{0}$~\cite{Rinke09} and previous HSE and DMC calculations~\cite{Leung99,Batista06}.

\section{Conclusion}

QMC methods present an accurate tool for the calculation of point defect formation energies, provided care is taken to control the accuracy of all the underlying approximations.  Including corrections for the approximations yields DMC values for the Si interstitial defects on par with $GW$ and hybrid-functional DFT calculations.  While backflow transformation and zero-variance extrapolation to remove the fixed-node error modify the energies slightly, further work remains to carefully control for finite-size effects known to plague defect supercell calculations.

\begin{acknowledgement}

The work was supported by the U.S. Department of Energy under Contract No.\ DE-FG02-99ER45795 and DE-FG05-08OR23339 and the National Science Foundation under Contract No.\ EAR-0703226.  This research used computational resources of the National Energy Research Scientific Computing Center, which is supported by the Office of Science of the U.S. Department of Energy under Contract No.\ DE-AC02-05CH11231, the National Center for Supercomputing Applications under grant DMR050036, the Ohio Supercomputing Center and the Computation Center for Nanotechnology Innovation at Rensselaer Polytechnic Institute.  We thank Cyrus Umrigar and Ann Mattsson for helpful discussion.

\end{acknowledgement}

%
\bibliographystyle{pss}
\bibliography{SiliconDefectQMCReview.bib}

\providecommand{\WileyBibTextsc}{}
\let\textsc\WileyBibTextsc
\providecommand{\othercit}{}
\providecommand{\jr}[1]{#1}
\providecommand{\etal}{~et~al.}


\begin{thebibliography}{[10]}

\othercit
\bibitem{Tilley08}
 \textsc{R.\,J.\,D. Tilley},
Defects in Solids (Wiley, 2008).


\bibitem{Fahey89}
 \textsc{P.\,M. Fahey},  \textsc{P.\,B. Griffin},  and  \textsc{J.\,D.
  Plummer},
 \jr{Rev. Mod. Phys.} \textbf{61}(2), 289--384 (1989).


\bibitem{Eaglesham94}
 \textsc{D.\,J. Eaglesham},  \textsc{P.\,A. Stolk},  \textsc{H.\,J. Gossmann},
  and  \textsc{J.\,M. Poate},
 \jr{Applied Physics Letters} \textbf{65}(18), 2305--2307 (1994).


\bibitem{Pike97}
 \textsc{L.\,M. Pike},  \textsc{Y.\,A. Chang},  and  \textsc{C.\,T.
  Liu},
 \jr{Acta Materialia} \textbf{45}(9), 3709 -- 3719 (1997).


\bibitem{Zhu99}
 \textsc{J.\,H. Zhu},  \textsc{L.\,M. Pike},  \textsc{C.\,T. Liu},  and
  \textsc{P.\,K. Liaw},
 \jr{Acta Materialia} \textbf{47}(7), 2003 -- 2018 (1999).


\bibitem{Otsuka99}
 \textsc{K.~Otsuka} and  \textsc{X.~Ren},
 \jr{Intermetallics} \textbf{7}(5), 511 -- 528 (1999).


\bibitem{Pelaz09}
 \textsc{L.~Pelaz},  \textsc{L.\,A. Marqu\'{e}s},  \textsc{M.~Aboy},
  \textsc{P.~L\'{o}pez},  and  \textsc{I.~Santos},
 \jr{Eur. Phys. J. B} \textbf{72}(3), 323--359 (2009).


\bibitem{Jones89}
 \textsc{R.\,O. Jones} and  \textsc{O.~Gunnarsson},
 \jr{Rev. Mod. Phys.} \textbf{61}(3), 689--746 (1989).


\bibitem{Ceperley80}
 \textsc{D.\,M. Ceperley} and  \textsc{B.\,J. Alder},
 \jr{Phys. Rev. Lett.} \textbf{45}(7), 566--569 (1980).


\bibitem{Perdew81}
 \textsc{J.\,P. Perdew} and  \textsc{A.~Zunger},
 \jr{Phys. Rev. B} \textbf{23}(10), 5048--5079 (1981).


\bibitem{PW92}
 \textsc{J.\,P. Perdew} and  \textsc{Y.~Wang},
 \jr{Phys. Rev. B} \textbf{45}(23), 13244--13249 (1992).


\othercit
\bibitem{PW91}
 \textsc{J.\,P. Perdew},
Unified {T}heory of {E}xchange and {C}orrelation {B}eyond the {L}ocal {D}ensity
  {A}pproximation,
 in: Electronic Structure of Solids '91, edited by P.~Ziesche and H.~Eschrig,
  (Akademie Verlag, Berlin, 1991),  pp.\,11--20.


\bibitem{PBE}
 \textsc{J.\,P. Perdew},  \textsc{K.~Burke},  and
  \textsc{M.~Ernzerhof},
 \jr{Phys. Rev. Lett.} \textbf{77}(18), 3865--3868 (1996).


\bibitem{AM05}
 \textsc{R.~Armiento} and  \textsc{A.\,E. Mattsson},
 \jr{Phys. Rev. B} \textbf{72}(8), 085108 (2005).


\bibitem{WC}
 \textsc{Z.~Wu} and  \textsc{R.\,E. Cohen},
 \jr{Phys. Rev. B} \textbf{73}(23), 235116 (2006).


\bibitem{PBESol}
 \textsc{J.\,P. Perdew},  \textsc{A.~Ruzsinszky},  \textsc{G.\,I. Csonka},
  \textsc{O.\,A. Vydrov},  \textsc{G.\,E. Scuseria},  \textsc{L.\,A.
  Constantin},  \textsc{X.~Zhou},  and  \textsc{K.~Burke},
 \jr{Phys. Rev. Lett.} \textbf{100}(13), 136406 (2008).


\bibitem{Heyd05}
 \textsc{J.~Heyd},  \textsc{J.\,E. Peralta},  \textsc{G.\,E. Scuseria},  and
  \textsc{R.\,L. Martin},
 \jr{J. Chem. Phys.} \textbf{123}(17), 174101 (2005).


\bibitem{Nieminen09}
 \textsc{R.\,M. Nieminen},
 \jr{Model. Sim. Mat. Sci. Eng.} \textbf{17}(8), 084001 (2009).


\bibitem{Becke93}
 \textsc{A.\,D. Becke},
 \jr{J. Chem. Phys.} \textbf{98}(7), 5648--5652 (1993).


\bibitem{Heyd03}
 \textsc{J.~Heyd},  \textsc{G.\,E. Scuseria},  and
  \textsc{M.~Ernzerhof},
 \jr{J. Chem. Phys.} \textbf{118}(18), 8207--8215 (2003).


\bibitem{Richie04}
 \textsc{D.\,A. Richie},  \textsc{J.~Kim},  \textsc{S.\,A. Barr},
  \textsc{K.\,R.\,A. Hazzard},  \textsc{R.~Hennig},  and  \textsc{J.\,W.
  Wilkins},
 \jr{Phys. Rev. Lett.} \textbf{92}(4), 045501 (2004).


\bibitem{Du07}
 \textsc{Y.\,A. Du},  \textsc{R.\,G. Hennig},  \textsc{T.\,J. Lenosky},  and
  \textsc{J.\,W. Wilkins},
 \jr{Eur. Phys. J. B} \textbf{57}(3), 229--234 (2007).


\bibitem{Vaidyanathan07}
 \textsc{R.~Vaidyanathan},  \textsc{M.\,Y.\,L. Jung},  and  \textsc{E.\,G.
  Seebauer},
 \jr{Phys. Rev. B} \textbf{75}, 195209 (2007).


\bibitem{Batista06}
 \textsc{E.\,R. Batista},  \textsc{J.~Heyd},  \textsc{R.\,G. Hennig},
  \textsc{B.\,P. Uberuaga},  \textsc{R.\,L. Martin},  \textsc{G.\,E. Scuseria},
   \textsc{C.\,J. Umrigar},  and  \textsc{J.\,W. Wilkins},
 \jr{Phys. Rev. B} \textbf{74}(12), 121102 (2006).


\bibitem{Rinke09}
 \textsc{P.~Rinke},  \textsc{A.~Janotti},  \textsc{M.~Scheffler},  and
  \textsc{C.\,G.\,V. de~Walle},
 \jr{Phys. Rev. Lett.} \textbf{102}(2), 026402 (2009).


\bibitem{Leung99}
 \textsc{W.\,K. Leung},  \textsc{R.\,J. Needs},  \textsc{G.~Rajagopal},
  \textsc{S.~Itoh},  and  \textsc{S.~Ihara},
 \jr{Phys. Rev. Lett.} \textbf{83}(12), 2351--2354 (1999).


\bibitem{Foulkes01}
 \textsc{W.\,M.\,C. Foulkes},  \textsc{L.~Mitas},  \textsc{R.\,J. Needs},  and
  \textsc{G.~Rajagopal},
 \jr{Rev. Mod. Phys.} \textbf{73}(1), 33--83 (2001),
Sections {V}. and {VI}. contrast {QMC} and {DFT} results. Section {X}.{E.}
  discusses scaling with computer time. Section {III}. introduces {VMC} and
  {DMC}.


\othercit
\bibitem{Needs06}
 \textsc{R.~Needs},
Quantum {M}onte {C}arlo {T}echniques and {D}efects in {S}emiconductors,
 in: Theory of Defects in Semiconductors, edited by D.~Drabold and
  S.~Estreicher, , Topics Appl. Physics Vol.\,104 (Springer Verlag, Berlin,
  Heidelberg, 2006),  p.\,141.


\bibitem{Esler08}
 \textsc{K.\,P. Esler},  \textsc{J.~Kim},  \textsc{D.\,M. Ceperley},
  \textsc{W.~Purwanto},  \textsc{E.\,J. Walter},  \textsc{H.~Krakauer},
  \textsc{S.~Zhang},  \textsc{P.\,R.\,C. Kent},  \textsc{R.\,G. Hennig},
  \textsc{C.~Umrigar},  \textsc{M.~Bajdich},  \textsc{J.~Kolorenc},
  \textsc{L.~Mitas},  and  \textsc{A.~Srinivasan},
 \jr{J. Phys. Conf. Ser.} \textbf{125}, 012057 (15pp) (2008).


\bibitem{Umrigar07}
 \textsc{C.\,J. Umrigar},  \textsc{J.~Toulouse},  \textsc{C.~Filippi},
  \textsc{S.~Sorella},  and  \textsc{R.\,G. Hennig},
 \jr{Phys. Rev. Lett} \textbf{98}(11), 110201 (2007).


\bibitem{Lopez-Rios06}
 \textsc{P.~{L\'{o}pez R\'{\i}os}},  \textsc{A.~Ma},  \textsc{N.\,D. Drummond},
   \textsc{M.\,D. Towler},  and  \textsc{R.\,J. Needs},
 \jr{Phys. Rev. E} \textbf{74}(6), 066701 (2006).


\othercit
\bibitem{Golub96}
 \textsc{G.\,H. Golub} and  \textsc{C.\,F.\,V. Loan},
Matrix Computations, 3rd edition,
 (The Johns Hopkins University Press, Baltimore, 1996), chap.~2,  p.\,51.


\bibitem{Umrigar88}
 \textsc{C.\,J. Umrigar},  \textsc{K.\,G. Wilson},  and  \textsc{J.\,W.
  Wilkins},
 \jr{Phys. Rev. Lett.} \textbf{60}(17), 1719--1722 (1988).


\bibitem{Anderson75}
 \textsc{J.\,B. Anderson},
 \jr{J. Chem. Phys.} \textbf{63}(4), 1499--1503 (1975).


\bibitem{Umrigar93}
 \textsc{C.\,J. Umrigar},  \textsc{M.\,P. Nightingale},  and  \textsc{K.\,J.
  Runge},
 \jr{J. Chem. Phys.} \textbf{99}(4), 2865--2890 (1993).


\bibitem{Alfe04b}
 \textsc{D.~Alf\`e},  \textsc{M.\,J. Gillan},  \textsc{M.\,D. Towler},  and
  \textsc{R.\,J. Needs},
 \jr{Phys. Rev. B} \textbf{70}(21), 214102 (2004).


\bibitem{Alfe04}
 \textsc{D.~Alf\`e} and  \textsc{M.\,J. Gillan},
 \jr{Phys. Rev. B} \textbf{70}(16), 161101 (2004).


\bibitem{Lin01}
 \textsc{C.~Lin},  \textsc{F.\,H. Zong},  and  \textsc{D.\,M. Ceperley},
 \jr{Phys. Rev. E} \textbf{64}(1), 016702 (2001).


\bibitem{Williamson97}
 \textsc{A.\,J. Williamson},  \textsc{G.~Rajagopal},  \textsc{R.\,J. Needs},
  \textsc{L.\,M. Fraser},  \textsc{W.\,M.\,C. Foulkes},  \textsc{Y.~Wang},  and
   \textsc{M.\,Y. Chou},
 \jr{Phys. Rev. B} \textbf{55}(8), R4851--R4854 (1997).


\bibitem{Ewald21}
 \textsc{P.\,P. Ewald},
 \jr{Annalen der Physik} \textbf{369}(3), 253--287 (1921).


\bibitem{Chiesa06}
 \textsc{S.~Chiesa},  \textsc{D.\,M. Ceperley},  \textsc{R.\,M. Martin},  and
  \textsc{M.~Holzmann},
 \jr{Phys. Rev. Lett.} \textbf{97}(7), 076404 (2006).


\bibitem{Kwee08}
 \textsc{H.~Kwee},  \textsc{S.~Zhang},  and  \textsc{H.~Krakauer},
 \jr{Phys. Rev. Lett.} \textbf{100}(12), 126404 (2008).


\bibitem{Drummond08}
 \textsc{N.\,D. Drummond},  \textsc{R.\,J. Needs},  \textsc{A.~Sorouri},  and
  \textsc{W.\,M.\,C. Foulkes},
 \jr{Phys. Rev. B} \textbf{78}(12), 125106 (2008).


\bibitem{Badinski10}
 \textsc{A.~Badinski},  \textsc{P.\,D. Haynes},  \textsc{J.\,R. Trail},  and
  \textsc{R.\,J. Needs},
 \jr{Journal of Physics: Condensed Matter} \textbf{22}(7), 074202 (2010).


\bibitem{Mitas91}
 \textsc{L.~Mit\'{a}\v{s}},  \textsc{E.\,L. Shirley},  and  \textsc{D.\,M.
  Ceperley},
 \jr{J. Chem. Phys.} \textbf{95}(5), 3467--3475 (1991).


\bibitem{Casula06}
 \textsc{M.~Casula},
 \jr{Phys. Rev B} \textbf{74}(16), 161102 (2006).


\bibitem{Pozzo08}
 \textsc{M.~Pozzo} and  \textsc{D.~Alf\`{e}},
 \jr{Phys. Rev. B} \textbf{77}(10), 104103 (2008).


\bibitem{Esler10}
 \textsc{K.\,P. Esler},  \textsc{R.\,E. Cohen},  \textsc{B.~Militzer},
  \textsc{J.~Kim},  \textsc{R.\,J. Needs},  and  \textsc{M.\,D. Towler},
 \jr{Phys. Rev. Lett.} \textbf{104}(18), 185702 (2010).


\bibitem{hood:076403}
 \textsc{R.\,Q. Hood},  \textsc{P.\,R.\,C. Kent},  \textsc{R.\,J. Needs},  and
  \textsc{P.\,R. Briddon},
 \jr{Phys. Rev. Lett.} \textbf{91}(7), 076403 (2003).


\bibitem{PhysRevB.71.035206}
 \textsc{J.~Shim},  \textsc{E.\,K. Lee},  \textsc{Y.\,J. Lee},  and
  \textsc{R.\,M. Nieminen},
 \jr{Phys. Rev. B} \textbf{71}(3), 035206 (2005).


\bibitem{PhysRevB.71.220101}
 \textsc{D.~Alf\`e} and  \textsc{M.\,J. Gillan},
 \jr{Phys. Rev. B} \textbf{71}(22), 220101 (2005).


\bibitem{PhysRevB.76.184103}
 \textsc{C.\,A. Gilbert},  \textsc{S.\,D. Kenny},  \textsc{R.~Smith},  and
  \textsc{E.~Sanville},
 \jr{Phys. Rev. B} \textbf{76}(18), 184103 (2007).


\bibitem{Bernholc88}
 \textsc{J.~Bernholc},  \textsc{A.~Antonelli},  \textsc{T.\,M. Del~Sole},
  \textsc{Y.~Bar-Yam},  and  \textsc{S.\,T. Pantelides},
 \jr{Phys. Rev. Lett.} \textbf{61}(23), 2689--2692 (1988).


\bibitem{Collins09}
 \textsc{A.\,T. Collins} and  \textsc{I.~Kiflawi},
 \jr{Journal of Physics: Condensed Matter} \textbf{21}(36), 364209 (2009).


\bibitem{Uberuaga05}
 \textsc{B.\,P. Uberuaga},  \textsc{R.~Smith},  \textsc{A.\,R. Cleave},
  \textsc{G.~Henkelman},  \textsc{R.\,W. Grimes},  \textsc{A.\,F. Voter},  and
  \textsc{K.\,E. Sickafus},
 \jr{Phys. Rev. B} \textbf{71}(10), 104102 (2005).


\bibitem{CASINO}
 \textsc{R.\,J. Needs},  \textsc{M.\,D. Towler},  \textsc{N.\,D. Drummond},
  and  \textsc{P.~{L\'{o}pez R\'{\i}os}},
 \jr{J. Phys. Cond. Matt.} \textbf{22}(2), 023201 (15pp) (2010).


\bibitem{QE-2009}
 \textsc{P.~Giannozzi},  \textsc{S.~Baroni},  \textsc{N.~Bonini},
  \textsc{M.~Calandra},  \textsc{R.~Car},  \textsc{C.~Cavazzoni},
  \textsc{D.~Ceresoli},  \textsc{G.\,L. Chiarotti},  \textsc{M.~Cococcioni},
  \textsc{I.~Dabo},  \textsc{A.\,D. Corso},  \textsc{S.~de~Gironcoli},
  \textsc{S.~Fabris},  \textsc{G.~Fratesi},  \textsc{R.~Gebauer},
  \textsc{U.~Gerstmann},  \textsc{C.~Gougoussis},  \textsc{A.~Kokalj},
  \textsc{M.~Lazzeri},  \textsc{L.~Martin-Samos},  \textsc{N.~Marzari},
  \textsc{F.~Mauri},  \textsc{R.~Mazzarello},  \textsc{S.~Paolini},
  \textsc{A.~Pasquarello},  \textsc{L.~Paulatto},  \textsc{C.~Sbraccia},
  \textsc{S.~Scandolo},  \textsc{G.~Sclauzero},  \textsc{A.\,P. Seitsonen},
  \textsc{A.~Smogunov},  \textsc{P.~Umari},  and  \textsc{R.\,M.
  Wentzcovitch},
 \jr{J. Phys. Cond. Matt.} \textbf{21}(39), 395502 (19pp) (2009).


\othercit
\bibitem{Wien2k}
 \textsc{P.~Blaha},  \textsc{K.~Schwarz},  \textsc{G.\,K.\,H. Madsen},
  \textsc{D.~Kvasnicka},  and  \textsc{J.~Luitz},
{WIEN2K}, {A}n {A}ugmented {P}lane {W}ave + {L}ocal {O}rbitals {P}rogram for
  {C}alculating {C}rystal {P}roperties ({K}arlheinz Schwarz, Techn.
  Universit\"{a}t Wien, Austria, 2001).


\bibitem{Trail05a}
 \textsc{J.\,R. Trail} and  \textsc{R.\,J. Needs},
 \jr{J. Chem. Phys.} \textbf{122}(1), 014112 (2005).


\bibitem{Trail05b}
 \textsc{J.\,R. Trail} and  \textsc{R.\,J. Needs},
 \jr{J. Chem. Phys.} \textbf{122}(17), 174109 (2005).


\othercit
\bibitem{CASINOpp}
 \textsc{J.\,R. Trail} and  \textsc{R.\,J. Needs},
{CASINO} pseudopotential library,
\url{http://www.tcm.phy.cam.ac.uk/~mdt26/casino2_pseudopotentials.html}.


\end{thebibliography}
 
\end{document}